# The effect of 20th century industrialization: Power station, acid rains, over-pumping, on an erstwhile uniform freshwater dune aquifer in Haifa Bay, Israel

Vasily Rogojin[1], Joel Kronfeld[2,*]

[1] Ministry of Environment, Ontario, Canada

[2] Department of Geophysics, School of Geosciences, Tel- Aviv University

*joel.kronfeld@gmail.com

**Abstract**

A small phreatic sand dune aquifer lies along the shore of Haifa Bay. It has been exploited for its freshwater resources since the 1930s. During this time the salinity has increased continuously, partly by sea water intrusion due to over pumping. The chemistry of the young aquifer water is laterally variable, and characterized by excess $SO_4^{2-}$, high $Sr^{2+}$ concentrations above that of modern sea water, high alkalinity, and markedly enriched $\delta^{13}C_{DIC}$ values. Acidic winter rains, formed from $SO_x$ and NOx gaseous emissions from a nearby power station, leach the dry deposition that had been added across the dune surface during the dry summers. The acidity also partially dissolves the aragonite sea shells in the dune sands, remainders of a previous marine transgression. As a consequence, this adds $Sr^{2+}$, $Ca^{2+}$-excess, and alkalinity, while leading to enriched $\delta^{13}C_{DIC}$ values, particularly during the winter, at which time the radiocarbon activity in the DIC is noted to decrease.

# 1. Introduction

Fresh water is an important and often limiting resource in Israel. When the quality of groundwater is degraded, it is necessary to investigated the causes that have been responsible for the degradation. This is the aim of the present study.

## 1.1 Haifa Bay

The Haifa Bay region is the industrial heartland of Israel which also includes agricultural settlements as well as a growing urban population. Along the coastline of Haifa Bay, between the cities of Haifa and Akko, is a narrow strip of sand dunes (Fig.1). This comprises one of the local aquifers that has been exploited for drinking water since the 1930s. Several hundred wells have been drilled, though only a small percent are still functioning. Over the past decades, tens of wells belonging to the Mekoroth Water Company have been closed due to the presence of unspecified pollutants. Today, all of the active wells are concentrated within the municipal boundaries of the cities of Qiriyat Khaim and Qiriyat Motskin (Fig.1). The salinity of groundwater in the first wells drilled did not exceed 80 mg/l $Cl^-$ (Loewengart, 1964). Today the salinity of the groundwater has risen to between 200-250 mg/l $Cl^-$ and even to 300-350 mg/l $Cl^-$. The groundwater facies is characterized as being of a Ca-Na-Mg-Cl-$HCO_3$ type. Unfortunately, no complete chemical analyses were made during the first decades of exploitation. Most of the wells that belong to the agricultural settlements, municipalities, and factories do not have chemical data. The archival analyses are often spotty and only report the $Cl^-$ content (Blake and Goldschmidt, 1947). It was only in the 1950s when the Mekoroth Water Company started to perform chemical analyses that complete reliable analyses started to be available in the region. When the rise in salinity was first noted, several hypotheses were put forth to explain it. These included the proposals of penetration of the nearby sea water (e.g., Kafri and Ecker, 1964; Bar-Yosef *et al*., 1997), the ascent of ancient brines (Mandel and Arad, 1958), relics from the Plio-Pleistocene transgression (Goldschmidt, 1970), and the addition of salts arising from sea spray (Loewengart, 1961). The present study was designed to employ isotopic methods along with chemical analyses to uncover the salinization factors and to trace the effects of a variety of environmental contaminants in degrading the quality of the underlying groundwater.

## 1.2 The Recent Dune Aquifer, the study area.

Sand dunes are situated along the sea shore to a maximum width of less than 3 km, along the coast of Haifa Bay lying between the cities of Haifa and Akko (Fig.1). The maximal height of the coastal sand dunes is 15 m. They run in a linear fashion along a SW/NE trend parallel to the coast. The dunes are composed of unconsolidated beach sediments, consisting of quartz sands, shell rubble and marine shells, and clays. The total area of the aquifer is 33 $km^2$ and the maximum thickness of the aquifer unit is 20 m. This aquifer is hydraulically isolated from other aquifer units, though it is in contact with the sea at its western extension. It overlies a consolidated sandy unit of Pleistocene age but is separated from it hydraulically by a 6 m thick clay layer. This clay layer is continuous throughout the study area and is an effective aquiclude. The recent sand dunes are permeable and readily receive recharge precipitation that falls directly upon them. Though the rainfall is between 500-700 mm/yr., the Haifa Bay region is considered to be semi-arid because the rain is restricted solely to the short winter season. No rain is expected to fall at other times of the year which is generally hot and dry. Unfortunately, no information was collected during the drilling of the well, as to the porosity, nor to the rate and effectiveness of the infiltration of the rain recharge. However, data is available on a similar recent coastal sandy dune aquifer to the south, at Ashdod. These parameters are expected be similar here. At Ashdod (Carmi et al., 2018, in press) the porosity ranges between 40-55%. The infiltration was determined to be very rapid and effective (Rimon et al., 2007). Thus, we can achieve a ball park estimate as to the rate of refilling, or the minimum rate of turnover, of the water supply. If we take the average value of precipitation as 600 mm/yr., infiltrating 20 m of sand that consists of 40- 50 % voids, then it would take on the order of only two decades to replenish the water in the aquifer, assuming minimal evaporative loss. Thus, the water in the relatively small Haifa bay dune aquifer, open to the sea along all of its length, was young and homogenous before exploitation. Within the area of the sampled wells a petroleum powered electricity generating utility was built in 1962/63. It was projected to consume approximately 200,000 tons of heavy fuel annually, which it was estimated should add 15,000 tons of $SO_3$ to the atmosphere yearly (Loewengart, 1964).

## 2.Sampling and Methods

Water samples were collected from pumping wells. The water temperature and pH were measured immediately at the well head. Bicarbonate ($HCO_3^-$) was measured in the field by titration. The major chemical constituents were analyzed at the Chemical Laboratory of the Faculty of Agriculture in Rehovot. The radiocarbon was analyzed at the Weizmann Institute of Science, Rehovot, on the $CO_2$ that was extracted from the groundwater. The $CO_2$ was converted to ethane which was measured in gas proportional counters. The results are expressed in per mill (‰) after normalizing for $\delta^{13}C$=-25‰. The statistical counting error is at the 1 σ level of confidence (Hoefs,1997). The stable hydrogen and oxygen isotopes were also analyzed at the Radiocarbon Laboratory at the Weizmann Institute of Science in Rehovot, by mass-spectrometry. The results are reported in conventional δ (‰) notation relative to the SMOW-V standard. The $\delta\ ^{18}O$ and $\delta\ ^{2}H$ are reported at a precision better than 0.2‰ and 1‰ respectively. The $\delta\ ^{13}C$ values of the total dissolved inorganic carbon are reported relative to the VPDB standard (Hoefs, 1997) with a precision of 0.2 ‰. The strontium and nitrogen isotopic analyses were performed at the CSIR, Pretoria, South Africa using a multi-collector VG354 thermal ionization mass-spectrometer following standard procedures. The measured isotopic ratios for $^{87}Sr/^{86}Sr$ were normalized before being reported, to 0.710250 for NIST 987 standard. The precision of the analytical results is taken (at 1 σ) as generally better than ± 0.000018. The $\delta^{15}N$ values are reported relative to atmospheric air with a precision of 0.15‰. The standards, the analytical methods, and the notation used are well explained in Hoefs (1997).

## 3. Results

The list of sampling wells, the thickness of the aquifer, the temperature, pH and conductivity of the ground water are presented in Table 1. The results of the chemical analyses of the groundwaters are presented in Table 2. Table 3a presents the results of the hydrogen, carbon and oxygen isotopic analyses of the studied groundwaters. The oxygen isotopic composition of the water falls in the range of from $\delta^{18}O$ = -4.1 ‰ to -4.4‰in the Blue Band wells (samples #1and #3) nearest to the shore line. The values rapidly become more depleted with distance from the sea (-4.7‰ to -5.1‰ in the Q. Motskin and Q. Khaim well fields). The deuterium varies in the same manner from $\delta^{2}H$ of -20.7‰ to -26.5 ‰ for the three samples for which deuterium analyses were performed. The relationship between $\delta^2H$ and $\delta^{18}O$ in the average local precipitation is defined by: $\delta^2H= 8\delta^{18}O +22$ (Gat and Dansgaard, 1972), termed the Eastern Mediterranean Water Line (EMWL). The groundwater in the dune aquifer shows a relationship of $\delta^2H= 8\delta^{18}O +14$. This deviation from the EMWL is explained by surface evaporation of the precipitation that affects the isotopic composition of the infiltrating water.

The $\delta^{13}C_{DIC}$ (= the isotopic composition of dissolved inorganic carbon) varies from -2.5‰ to -9.2 ‰. This is more enriched in $^{13}C$ than would be expected of groundwater receiving $CO_2$ from $C_3$ plants. Investigations of the unsaturated zone in the semi-arid regions of Israel have shown that the carbon isotopes in soil- gas $CO_2$ derived from $C_3$ trees and seasonal grasses, change from dissolved $CO_2$ ($\delta^{13}C$ = -22 ‰) to $HCO_3^-$ ($\delta^{13}C$ = -14 ‰) (Carmi et al, 2013; Carmi et al., 2015). Deeper within the unsaturated zone other processes change the $\delta^{13}C_{DIC}$ leading to the groundwater attaining an average $\delta^{13}C_{DIC}$ of -12.5 ‰. The Haifa bay region having a Mediterranean climate is exclusively dominated by $C_3$ plants (Vogel et al., 1986; Frumkin et al., 2000). The $\delta^{13}C_{DIC}$ in the dune aquifer is not only much more enriched but it is highly variable from well to well, as will be explained.

The radiocarbon concentrations in the DIC of the groundwater vary from approximately $\Delta^{14}C$= -194‰ to $\Delta^{14}C$= -454‰ for the five relatively closely spaced water wells. The water in these wells has infiltrated rapidly and should thus be young. It is known that the early Holocene sea level rose to a maximum height approximately 4000 years ago, transgressing over the coastal plain study area (Porat et al., 2008). As the sea withdrew over the next centuries it left behind shallow marine sediments including coarse sand containing plentiful mollusk shell material, (similar to those in the sediments at the present day foreshore of Haifa Bay). Shell material, was separated from the dune sediments and the $\delta^{13}C$ and $^{14}C$ activity measured. The carbon isotopic analyses are reported in Table 3b. The $\delta^{13}C$ of marine carbonate is close to 0‰. The sea shells exhibit $\delta^{13}C$ values consistent with a marine origin. The radiocarbon values of the shells range from $\Delta^{14}C$ = -728‰ to $\Delta^{14}C$ = -874‰. These non-modern values also are consistent with shells derived from the previous marine transgression.

The $Sr^{2+}$ concentration and $^{87}Sr/^{86}Sr$ ratios are presented in Table 4. The $Sr^{2+}$ concentrations fall in the range of 2.2 to 12 mg/l $Sr^{2+}$. This would place it among the high values for fresh water. Ocean water contains 8 mg/l $Sr^{2+}$ which is generally higher than fresh water (Hem, 1985). The $^{87}Sr/^{86}Sr$ ratios are consistent with modern sea water values (McArthur, *et al*., 2012).

The $\delta^{15}N$ of the $NO_3$ is presented in Table 4. The $NO_2^-$ is at or below the analytical detection limits, except for a value of 1.3 mg/l $NO_2^-$ in the Blue Band-3 well. Otherwise, all of the nitrogen resides as $NO_3^-$. The $\delta^{15}N$ of the N of the $NO_3^-$ ranges from +10 to + 18‰. The nitrogen concentrations vary from 11 mg/l to 25 mg/l $NO_3^-$.

## 4. Discussion.

### 4.1 Chemistry

The phreatic dune aquifer, lying in direct contact to the sea to the west is isolated from the underlying aquifer. It can receive salts from the sea either by sea spray or by direct intrusion. As a first step to unraveling the underlying causes for the measured chemical and isotopic deviations from what would be expected *a priori*, the possible contributions of the nearby sea has to be examined (Bar- Yosef et al., 1997; Arad et al., 1975). The sea-spray would add salts in a rather uniform amount yearly. This would have established a salinity baseline since the aquifer has been formed. However, sea-spray alone would not explain the increase in salinity that has been noted during recent decades by pumping.

To evaluate whether or not direct sea water intrusion is the dominant salinity source, it is necessary to calculate if sea water additions were added to an original uncontaminated water source to yield the ionic concentrations that were actually measured in the aquifer's waters (Table 2). As has been pointed out, no complete chemical analyses are available from the time before the waters started to experience salinization. To overcome this lacuna, the chemical analysis of the freshest water presently within the regional Haifa Bay catchment area is used. The freshest waters, of Ca- $HCO_3$ type, can be found in the replenishment area of the Megged Kramim well field (all of the wells there yield water of very similar chemistries). Megged Kramim -3 (sample #43 in Fig.1 and Table 2) is located approximately 26 km to the east, in the Galilean Hills, far from industrial contamination sites. It will be taken to represent an initial chemically "pure" starting water to which a fraction of sea water is to be conceptually added. Due to high net evaporation over the Mediterranean Sea the sea water salinity varies seasonally, with depth, and geographically (e.g. Nessim et al., 2015). Thus, an average value of 3.86% salinity (chloride = 21,200 mg/l) was used as representative for the Eastern Mediterranean (Lenntech, 2018). The addition of each ionic component, at each well, was computed according to the simple equation:

(1) $I_{sample} = I_{fresh\ end\ member} + I_{sea\ water}$,

Where $I_{sample}$ is the ion concentration (for $Ca^{2+}$, $Mg^{2+}$, $Na^{-}$, $SO_4^{2-}$, calculated sequentially), in mg/l, of the water sample at each well. It should yield the theoretically expected concentration for each ion considered as a result of mixing between the recharge water, $I_{fresh\ end\ member}$ and sea water, $I_{sea\ water}$. The sea water's ionic components are computed, normalized to the chloride content. The chloride ion was chosen for it behaves conservatively. It does not precipitate. It is not a component of the mineral structure of the carbonate nor is there any evaporate minerals in the aquifer. It does not participate in any known ion-exchange reactions. This equation normalizes the cations and anions in each sample against the concentration of chloride in modern Mediterranean Sea water. This fraction was evaluated as potentially representing its proportion in sea water in the manner:

(2) $I_{added} = (I_{sea\ water} / Cl^-_{sea\ water}) \times (Cl^-_{sample} - Cl^-_{fresh\ end\text{-}member})$,

where I is the concentration of the ion studied. Using the Q. Motskin-4a well as an illustrative example, the values of the measured ions were plugged into the equation and yield:

(3) $Ca^{2+}$ mixed = 80 + (423/21,200) x (177-25) = 80 + 3.0 = 83.0(mg/l)

Where: 80 is the concentration of $Ca^{2+}$ (mg/l), in the recharge along with 25 mg/l $Cl^-$ (represented by the Megged Kramin 3 well (Table 2); 423/21,200 is the normalization factor (the ratio of $Ca^{2+}/Cl^-$ in Mediterranean Sea water), 177 (in mg/l) is the concentration of $Cl^-$ that had been measured in the Q. Motskin 4a well. Thus 177-25 should be the amount of sea water contribution to the mixture for this sample.

Likewise, the other components can be calculated, using the measured ionic data for this well (Table 2), with the "original" fresh recharge components represented by Meggid Kramim 3 (Table 2), thus:

(4) $Mg^{2+}$ mixed = 35.1 + (1403/21200) x (177-25) = 45.2 (mg/l)

(5) $Na^+$ mixed = 13.0 + (11800/21200) x ( 177-25) = 97.6 (mg/l)

(6) $SO_4^{2-}$ mixed = 8.0 + (2950/21200) x (177-25) = 29.2 (mg/l)

For the Q. Motskin 4a well, these would be the expected ionic concentrations of the groundwater if sea water had been the only end member mixing with the starting "pure" sample. However, these values differ from the actual measured ion concentrations ($Ca^{2+}$ = 96 mg/l; $Mg^{2+}$ =34 mg/l; $Na^{+}$ = 103 mg/l; $SO_4^{2-}$ = 73 mg/l). The extent of these deviations can best be seen by the difference, Δ, between the ions in the samples to that hypothetically evolved from mixing with modern Mediterranean Sea water (Table 5). This calculation was carried out for all of the well waters. There is a marked excess of $Na^{+}$, $Ca^{2+}$, and $SO_4^{2-}$ compared to sea water, which can be as great as 130-140 mg/l (Table 5). Neither sea water nor sea spray is supplying this additional amount of ionic components. However, there are anthropogenic sources in the vicinity that spew out large quantities of pollutants. These include the power station, a fertilizer plant and building material factories. It was at first suspected that the dominant factor is the power station which releases extensive amounts of NOx and $SO_3$. The dune system is downwind of the power plant. In the rainless summer season a dry deposition occurs over the surface of the dunes (Loewengart,1964). During the winter months the NOx and $SO_3$ gases combine with the precipitation to form acid rain, ($HNO_3$ and $H_2SO_4$). The acidic precipitation (pH as low as 4, though averaging 6 [Mamane, 1987a, b]) dissolves the dry-deposited pollutants. The effects of the acidic precipitation upon the soil will depend upon the amount of rain and its acidity at a specific location as well as the acid soluble proportion of the deposited fly ash and the amount of shells lying exposed on the dunes. The dune aquifer is characterized by a large and variable excess of $SO_4^{2-}$ in particular that can be explained as coming from acid rains. The "excess –$SO_4$" can be used to signify the intensity of the acidity that washes through the dunes.

### 4.2 Isotopes

The acidity as it percolates through the sands not only adds $SO_4^{2-}$, but changes the carbon isotopic composition of the underlying aquifer. The acidity readily interacts with the carbonate shells which neutralize it. By doing so, several important isotopic changes will transpire. Normally, root respiration and decay of plants will form bicarbonate with a $\delta^{13}C$ of ~ -14‰ (under a $C_3$ plant regime) having radiocarbon values similar to atmospheric values. In the present situation the acidity dissolves carbonate shells. The addition of $CO_2$ released from the dissolving marine carbonates ($\delta^{13}C$ = ~0‰) will form $\delta^{13}C_{DIC}$ of some intermediate values, depending upon the relative amounts of the two $CO_2$ components. This is demonstrated in the five seasonal pairings of $\delta^{13}C$ values measured both in summer and winter in the DIC of the Q. Motskin 4a,12, and 18 wells and two pairs measured in the Q. Khaim 14 well (two pairs) (Table 3a). The winter values are always enriched compared to the summer analyses, as this is the time when the acid rain dissolution of the marine shells occurs. This explains the variable and enriched $\delta^{13}C_{DIC}$ values. Likewise, the addition of $CO_2$ gas released from the older carbonate marine shells (Table 3b) will add alkalinity and affect the radiocarbon activity in the DIC of the water. In the Q. Motskin-12 well the $^{14}C$ activity was measured once in winter and once in summer. The winter analysis, yields a lower $^{14}C$ activity than that of the summer (Table 3a). Assuming that the winter result (Table 3b) is due to dissolution of older marine shells from the last marine transgression by acid rains, a mass balance calculation between the $^{14}C$ and the $\delta^{13}C$ of the water and shells was performed (appendix 1). The water from the Q. Motskin-4a well (Table 3a) was used as one end member and shells from near the power stations and next to Q. Haim-4 were used as the other end member. It can be calculated that the winter DIC in the Q. Motskin- 12 well water is derived from 78 to 83% of the DIC in the residual summer aquifer water and between 17 to 22% from dissolved shells (appendix 1).

The partial dissolution of the marine shells, admixed in the dune sands, explains the uniform (0.7090-0.7091) $^{87}Sr/^{86}Sr$ ratios reflective of modern sea water (MacArthur et al., 2012). The high $Sr^{2+}$ concentrations coupled with the modern sea water isotopic ratios can be explained by the transformation of the aragonite marine shells releasing up to 10 molar percent of $Sr^{2+}$ that is contained in the aragonite (Morse and Mackenzie, 1990) as it reverts to the more stable calcite mineral form. This mineralogical transformation can occur rapidly in the presence of fresh water (percolating rain recharge). The release of $Sr^{2+}$ from aragonite is would also be assisted by leaching by the acid rains The $Sr^{2+}$ derived from the marine aragonite would subsequently be added into the groundwater.

The $\delta^2H$ and the $\delta^{18}O$ of the water are neither affected by acidity nor by emissions from the power plant. Their isotopes remain similar to the local precipitation, modified slightly by surficial evaporation during recharge Though the oxygen isotopic value of sea water is much more enriched than the local precipitation, only a small contribution of sea water (21,200 mg/l $Cl^-$) to the initial groundwater (<80 mg/l $Cl^-$) is needed in order to significantly change its chlorinity. This amount of mixing, though it affects the water quality, would not affect the $\delta^{18}O$ of the resulting mixture for the chlorinity values encountered here.

The $NO_3^-$ might be expected be related to the emissions from the power station. Acid rain can add $NO_3^-$ to the ground water from the percentage of the $HNO_3$ in the acidity that makes up the acid rain. The nitrate concentrations in the wells studied are variable both between wells and between years from the same well. All values are within the EPA limits, and within the concentration range encountered in the fresh underlying aquifers of the region (Rogojin et al., 2002). However, the dune aquifer waters exhibit $\delta^{15}N_{NO3}$ that are distinctly enriched compared to groundwater nitrates derived from soil inorganic nitrate, whose $\delta^{15}N$ values should range of +2

to+9 ‰ (e.g. Gormly and Spalding, 1979, [ $\delta^{15}N$ of +5 to +9‰ ]; Kreitler et al., 1978 [$\delta^{15}N$ =+2 to +8‰) ]). The $\delta^{15}N$ values in the dune aquifer are much more enriched. The $\delta^{15}N$ isotopic values fall within the range of +10 to +18‰. There are several potential ways that can lead to the high isotopic enrichments, including denitrification, deposition from coal based power plants and to catabolic reactions manifest in agriculture organic fertilizer (manure) or sewage (Heaton, 1986). Livestock wastes have been reported to range from +10‰ to +20‰ and higher (e.g. Gormly and Spalding, 1979; Kreitler et al., 1978). As the aquifer is highly porous and oxygenated, de-nitrification does not take place; for, it is an anoxic reaction. The emissions and their subsequent particulate deposition are from a petroleum-based power station whose emissions should be close to zero or slightly negative (Heaton, 1990). As neither agriculture nor livestock pens have been present within the municipal boundaries of Q. Motskin nor in Q. Khaim, animal sources are not the cause. Thus, sewage leakage seems the most plausible, though not attractive explanation, to explain the nitrate sources in the wells analyzed that supplied drinking water in the municipalities. It may well be that the values encountered in the wells are a mixture of sources, such as that of power station emissions (relatively depleted) combined with the sewage values (highly enriched) to yield the intermediate $\delta^{15}N$ isotopic values reported here in the municipal supplies. Indeed, in Q. Motskin, it was recommended to close two municipal wells due to high values of “organic matter”, noted to have been present (Ministry of Health, reported to the Israel Water Commission,1998). The elevated value of $\delta^{15}N$ = +10 ‰ noted in the water of the Blue Band Factory well is most probably not related to sewage as are the other wells. This factory produces dairy products, and the slightly elevated $\delta^{15}N$ is probably related to this source.

## 5.Conclusions

The porosity is high and the infiltration in to the unconsolidated dunes is rapid. The recent dune aquifer is small, yet, the shape is narrow and long, hugging the coast. The mixing therefore would not be rapid, along its entirety; but would allow for contaminant point sources (at least over the short term) to maintain differences among wells. The aquifer is hydraulically isolated from underlying aquifers. Laterally sea water intrusion from the west is likely through over pumping. The only way that ions not related to sea water can enter is through the top of the aquifer and then percolate downwards. The acidic emissions from the nearby power plant are point sources of acidity along the aquifer surface. Local inputs are rapid and the chemical and isotopic inputs vary significantly from place to place and over time. Pollutants can degrade the water quality as has been attested to be on going the since the 1930s. The acid emissions are the dominant factor that changes the water chemistry and isotopic composition of the local dune sources above that of the sea – sourced contributions. The power station emissions are wind borne and winds shift direction. Therefore, the pollutants can be distributed non-homogenously over the dunes. Thus geographically variable chemical and isotopic signatures denoted in the groundwater sites can be expected to change over time.

The power station is now completing its conversion from petroleum to natural gas. This should significantly reduce the NOx and $SO_3$ gas emissions (the acidity) and also the dry deposition of fine ash. If pumping is not increased or is even decreased the isotopic and chemical signature of the ground water of the entire aquifer will change, homogeneously, in the direction of the pre-exploitation chemical state. The isotopic and chemical composition in the future would thus be influenced dominantly by that of the local precipitation and sea spray. Any salinity increases would be dependent upon the degree of aquifer exploitation (affecting the degree of sea water intrusions). This study of a small isolated coastal aquifer, being affected by sea water urbanization, and industrialization, can be used as a model to explain contamination processes that might occur in other coastal aquifers.

**Appendix: Dissolution of shells in winter by acid rains and the effect on the aquifer water**

Dissolution of shells by acid rain in the winter and the effect on the carbon isotopes in the DIC of the aquifer is calculated by a mass balance approach using the carbon isotopic data in Table 3a. The end members are the $^{14}C$ in the Q. Motskin- 12 well (Table 3a) and the sea shells near the Q. Khaim wells (Table 3b).

The $^{14}C$ values of Motzkin-12 are: 56.7 pMC in the winter and 65.8 pMC in the summer. The mixing of the component derived from acid rain dissolution of shells in the winter with the residual summer ground water values should equal the groundwater $^{14}C$ value measured in the winter. To calculate the percentage contribution of the shells by acid rain dissolution and mixing with the residual summer $^{14}C$ value of the $^{14}C$ activity in the dune shells is 28.6 pMC and for the shells at Q. Khaim (23.3 pMC and 13.3 pMC) to get the $^{14}C$ winter value of 56.7 pMC for Motzkin-12:

1. 56.7=65.8X+23.3x (1-) yields mixing contributions of 78.6% groundwater DIC and 21.4% from shells
2. 56.7=65.5X+13.3x (1-X) yields mixing contributions of 82.7% DIC and 17.3% from the shells

## Highlights

- local power plant releases NOx and $SO_3$ dry deposition during the summer over the aquifer sands.
- Winter rains raise the surface acidity that dissolves the admixed sea shells adding sulfate and strontium to the groundwater.
- Over-pumping has enabled sea water intrusion to raise the chlorinity

- Sewage inputs have been denoted using nitrogen isotopes
- The change to natural gas in the power station should reduce a significant component of pollution going forward.

## Acknowledgements

This study was supported by Tel Aviv University Research Funds. The authors would like to thank Mr. Siep Talma, of the CSIR (Pretoria) for the analyses of the nitrogen isotopes and for two replicate analyses of $\delta^{13}C$ and $\delta^{18}O$. Dr. Bruce Eglington is thanked for measuring the isotopes of strontium. The authors would like to thank an anonymous reviewer for improving the manuscript.

| Conductivity µS/cm | pH | T ( °C) | Depth (m) | Well | Sample |
|---|---|---|---|---|---|
| 1620 | 7.55 | 24.0 | 19.0 | Blue Band – 3 | 1 |
| 1375 | 7.40 | 23.5 | 19.7 | Blue Band Factory | 3 |
| 1190 | 7.56 | 23.0 | 19.0 | Q.Motskin – 4a | 25 |
| 1220 | 7.33 | 23.8 | 13.9 | Q.Motskin- 12 | 28 |
| 1270 | 7.38 | 23.3 | 14.4 | Q.Motskin -18 | 29 |
| 1590 | 7.44 | 23.0 | 17.8 | Q.Khaim- 14 | 35 |
| 1710 | 7.40 | 23.5 | 19.5 | Q.Khaim -6 | 30 |
| 670 | 7.40 | 23.3 | 540 | Megged Kramim-3 | 43 |

Table1. A list of sampled wells from the recent dune aquifer with the sample location code ( Fig.1), the depth of the well, as well as the temperature, pH, and conductivity of the groundwater.

| $HCO_3^-$ mg/l | $SO_4^{2-}$ mg/l | $Cl^-$ mg/l | $Na^+$ mg/l | $Mg^{2+}$ mg/l | $Ca^{2+}$ mg/l | $K^+$ mg/l | Date of analysis | Well |
|---|---|---|---|---|---|---|---|---|
| 388.0 | 82.0 | 208.0 | 170.0 | 34.1 | 98.5 | 8.3 | 28.07.92 | Blue Band-3 |
| 317.0 | 88.0 | 149.0 | 100.0 | 31.4 | 100.0 | 5.8 | 28.07.92 | Blue Band Factory |
| 338.9 | 73.0 | 177.0 | 103.0 | 33.9 | 96.0 | 6.2 | 18.12.95 | Q. Motskin-4a |
| 329.5 | 168.1 | 209.2 | 131.0 | 46.2 | 114.2 | 2.7 | 07.08.74 | Q. Motskin-12 |
| 329.5 | 124.9 | 173.7 | 96.6 | 42.6 | 98.2 | 2.0 | 07.08.74 | Q. Motskin-18 |
| 396.6 | 158.5 | 202.1 | 131.0 | 60.8 | 96.2 | 9.0 | 07.08.74 | Q. Haim-14 |
| 353.9 | 105.7 | 234.0 | 137.9 | 49.9 | 110.2 | 18.4 | 08.09.73 | Q. Haim-6 |
| 395.0 | 8.0 | 25.0 | 13.0 | 35.1 | 80.0 | 1.3 | 15.07.94 | Megged Kramim-3 |

Table 2. Water chemistry of the studied aquifer. The Megged Kramim-3 well is not in the studied aquifer: it represents the freshest and least altered groundwater within the watershed. It is used for calculations in the paper as a close representative of the "original" groundwater prior to pollution.

| Sample | Well | Date | $\delta^2H$ (‰) | $\delta^{18}O$ (‰) | $\delta^{13}C$ (‰) | $^{14}C$ (pMC) | $\Delta^{14}C$ (‰) |
|---|---|---|---|---|---|---|---|
| 1 | Blue Band–3 | 28.07.92 | -20.7 | -4.4 | | | |
| 3 | Blue Band Factory | 28.07.92 | | -4.1 | | | |
| 25 | Q. Motskin–4a | 18.12.95 | | | -5.0 | | |
| | | 15.06.96 | | -4.7 | -7.9 | 65.5±0.3 | -367±3 |
| | | 15.06.97 | | | -7.9 | 63.5±0.4 | -386±4 |
| 28 | Q. Motskin-12 | 18.12.95 | | -4.8 | | 56.7±0.3 | |
| | | 15.06.97 | | | -6.3 | 65.8±0.3 | -367±3 |
| | | 13.01.99 | | -4.4 | -5.0 | | |
| 29 | Q.Motskin-18 | 20.01.96 | | | -7.3 | | |
| | | 15.06.97 | | -4.9 | -9.2 | 73.6±0.4 | -287±4 |
| | | 13.01.99 | | -4.6 | -7.3 | | |
| 35 | Q.Khaim-14 | 20.01.96 | | | -2.5 | | |
| | | 15.06.97 | -26.5 | | -7.3 | 83.5±0.3 | -194±3 |
| | | 29.06.98 | | -5.1 | -7.4 | | |
| | | 13.01.99 | | -5.0 | -6.8 | | |
| 30 | Q.Khaim-6 | 18.12.95 | -26.0 | -5.0 | -5.9 | 68.2±0.3 | -344±3 |

Table 3a. The hydrogen, oxygen, and carbon isotopic data from the sampled groundwaters.

| Collection site | Depth (m) | $\delta^{13}C$ (‰) | $^{14}C$ (pMC) | $\Delta^{14}C$ (‰) |
|---|---|---|---|---|
| Dune 180m from power station | surface | +0.1 | 28.6±0.2 | -728±2 |
| Next to Q. Khaim - 14 | 0.1 – 0.3 | -0.3 | 13.3±0.1 | -874±1 |
| Next to Q. Khaim - 6 | 0.2 – 0.5 | -1.0 | 23.3±0.5 | -778±5 |

Table 3b. The carbon Isotopic values from sea shells admixed with the dune sediments.

| $\delta^{15}N$ (‰) | $NO_2^-$ mg/l | $NO_3^-$ mg/l | $^{87}Sr/^{86}Sr$ | $Sr^{2+}$ mg/l | Date of analysis | Well |
|---|---|---|---|---|---|---|
| 10.5 | 1.3 | 31.4 | 0.70900 | 2.2 | 28.07.92 | Blue Band-3 |
| 10.5 | | 7.1 | | | 20.08.94 | Blue Band-3 |
| | | | 0.70910 | 2.6 | 28.07.92 | Blue Band Factory |
| 13.4 | | 2.7 | | | 17.08.94 | Q. Motskin-4a |
| 13.4 | <0.3 | 12.0 | | | 18.12.95 | Q. Motskin-4a |
| | <0.3 | 13.7 | 0.70899 | 6.1 | 07.08.74 | Q. Motskin-12 |
| 15.0 | | 3.1 | | | 24.08.94 | Q. Motskin-12 |
| | | | | | 07.08.74 | Q. Motskin-18 |
| 18.4 | <0.3 | 16.4 | 0.70903 | 6.0 | 07.08.74 | Q. Haim-14 |
| 18.4 | | 3.7 | | | 24.08.94 | Q. Haim-14 |
| 18.2 | <0.3 | 25.5 | 0.70914 | 12.9 | 08.09.73 | Q. Haim-6 |
| 18.2 | | 5.7 | | | 24.08.94 | Q. Haim-6 |

Table 4. The strontium concentrations and isotopic ratios in the studied groundwaters, as well as the nitrate concentration and associated nitrogen isotopic composition in these groundwaters.

| measured | calculated | Δ (measured-calculated) | | | | | | | | | |
|---|---|---|---|---|---|---|---|---|---|---|---|
| Well | $Ca^{2+}$ mg/l | $Na^{+}$ mg/l | $SO_4^{2-}$ mg/l | $Ca^{2+}$ mg/l | $Mg^{2+}$ mg/l | $Na^{+}$ mg/l | $SO_4^{2-}$ mg/l | $Ca^{2+}$ mg/l | $Mg^{2+}$ mg/l | $Na^{+}$ mg/l | $SO_4^{2-}$ mg/l |
| Q. Motskin-4a | 96.0 | 103.0 | 76.0 | 83.0 | 45.2 | 97.6 | 29.2 | 13.0 | -11.2 | 5.4 | 43.8 |
| Q. Motskin-12 | 114.2 | 131.0 | 168.1 | 83.7 | 47.3 | 115.4 | 33.6 | 30.7 | -1.1 | 15.6 | 134.5 |
| Q. Motskin-18 | 98.2 | 96.6 | 124.9 | 83.0 | 44.9 | 95.8 | 28.6 | 25.2 | -2.3 | 0.8 | 96.3 |
| Q. Haim-14 | 108.2 | 119.6 | 81.7 | 84.0 | 48.2 | 123.4 | 35.6 | 24.2 | -5.6 | -3.8 | 46.1 |
| Q. Haim-6 | 110.2 | 137.9 | 105.7 | 84.2 | 48.9 | 129.3 | 37.1 | 26.0 | 1.0 | 8.6 | 68.6 |
| Blue Band-3 | 98.5 | 170.0 | 82.0 | 83.7 | 47.2 | 114.9 | 33.5 | 14.8 | -13.1 | 55.1 | 48.5 |
| Blue Band Factory | 100.0 | 100.0 | 88.0 | 82.5 | 43.3 | 82.0 | 25.3 | 17.5 | -11.9 | 18.0 | 62.7 |

Table 5. Calculating what the chemical composition of the groundwater should be if the increase in salinity (starting with the freshest water in the water shed, the Megged Kramim 3 well) was solely due to the intrusion of Mediterranean seawater. The deviation, "Δ", is the difference from the groundwater values measured minus those values calculated. This is described fully within the text. Small deviations (both positive and negative) should be expected as only an average Mediterranean Sea water value was used, and as noted in the text there are variations in the sea water throughout the year due primarily to evaporation. Large differences though imply that another source(s), other than sea water, is/are responsible for the ionic values that have been measured in this study.

□

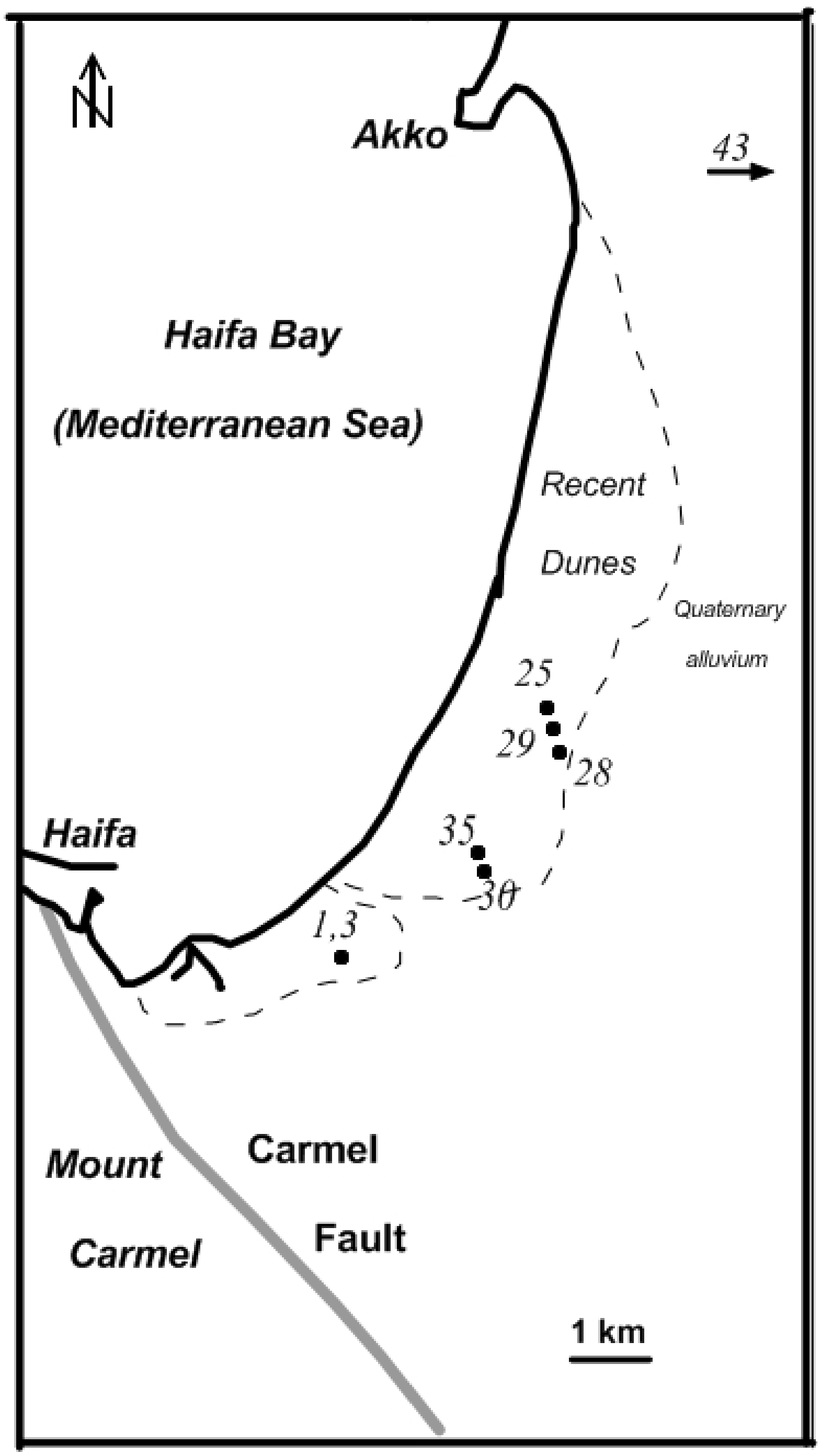

Figure 1. Sample location map, showing the sampled wells and the extent of the recent coastal dune aquifer along the margins of Haifa Bay. Sample #43 represents the Megged Kramim 3 well, far from pollution sources, having the freshest water within the watershed.